# Mathematical Model of Volume Kinetics and Renal Function after Burn Injury and Resuscitation

Ghazal ArabiDarrehDor[1], Ali Tivay[1], Ramin Bighamian[2], Chris Meador[3], George C. Kramer[3,4] Jin-Oh Hahn[1†], Jose Salinas[5]

[1]: Department of Mechanical Engineering, University of Maryland, [2]: U.S. Food and Drug Administration, [3]: Arcos, Inc. [4]: Department of Anesthesiology, University of Texas Medical Branch, [5]: U.S. Army Institute of Surgical Research, [†]: Corresponding Author

## DISCLAIMER

## ABSTRACT

This paper presents a mathematical model of blood volume kinetics and renal function in response to burn injury and resuscitation, which is applicable to the development and non-clinical testing of burn resuscitation protocols and algorithms. Prior mathematical models of burn injury and resuscitation are not ideally suited to such applications due to their limited credibility in predicting blood volume and urinary output observed in wide-ranging burn patients as well as in incorporating contemporary knowledge of burn pathophysiology. Our mathematical model consists of an established multi-compartmental model of blood volume kinetics, a hybrid mechanistic-phenomenological model of renal function, and novel lumped-parameter models of burn-induced perturbations in volume kinetics and renal function equipped with contemporary knowledge on burn-related physiology and pathophysiology. Using the dataset collected from 16 sheep, we showed that our mathematical model can be characterized with physiologically plausible parameter values to accurately predict blood volume kinetic and renal function responses to burn injury and resuscitation on an individual basis against a wide range of pathophysiological variability. Pending validation in humans, our mathematical model may serve as an effective basis for in-depth understanding of complex burn-induced volume kinetic and renal function responses as well as development and non-clinical testing of burn resuscitation protocols and algorithms.

## INTRODUCTION

Burn resuscitation is a major critical care challenge. Immediately post burn, a large amount of intravascular fluid and protein is shifted into the interstitial space due to a chain of events initiated by the heat-induced inflammatory response of the body [1], [2]. Subsequent plasma loss in the intravascular space compromises cardiac performance, which may lead to end organ hypo-perfusion, ischemia, and death in severe burns [3], [4]. In critical care units, this loss is compensated for by fluid replacement therapies to replenish blood volume (BV), guided by available burn resuscitation guidelines [5].

In today's clinical practice, physicians frequently adjust fluid resuscitation regimen to maintain an adequate urinary output (UO) as an indicator of fluid replacement [6]. But, there is a lack of consensus on the optimal burn resuscitation strategy, in terms of the timing, amount, and type of fluids to be administered [7]–[10]. Considering that insufficient and delayed resuscitation may increase the mortality risk of burn patients [11], prior therapy in burn units tended to conservatively over-resuscitate patients with excessive amount of fluids up to twice as much as recommended [12], [13], exposing these patients to an elevated risk of side effects, e.g., pulmonary edema, limb and abdominal syndromes, necrosis, and death [14], [15] (known as fluid creep). Hence, optimizing burn resuscitation regimen may significantly contribute in reconciling the maintenance of organ function and the minimization of adverse complications. Conventional evidence-based approach to such optimization involves rigorous in vivo trials in animals and humans. In recent years, both the U.S. and European regulatory agencies have expressed interest in the use of mathematical models of physiological systems as a powerful non-clinical tool for developing and testing clinical therapies [16]–[19]. Hence, a credible mathematical model capable of predicting patient's physiological responses (including BV and UO) to burn injury and resuscitation has the potential to streamline the development of new optimal burn resuscitation regimen.

Prior work on mathematical modeling of burn injury and resuscitation exists. Arturson et al. [20]–[22] developed a phenomenological model using datasets collected from a small number of burn patients. Roa et al. [23]–[25] developed a hybrid mechanistic-phenomenological model using datasets collected from a number of burn patients. Bert et al. [26]–[30] also developed a hybrid mechanistic-phenomenological model using datasets collected from rats and humans. These models provide very good insights into the complicated physiology and pathophysiology associated with the burn injury and resuscitation. But at the same time, opportunities exist for improving the validity and versatility of existing mathematical models, especially in the context of testing burn resuscitation protocols and algorithms. First, a subset of these mathematical models cannot predict UO [26]–[30]. Such a limitation directly disqualifies a mathematical model in the development and testing of burn resuscitation regimen based on UO feedback. Second, the ability of the existing mathematical models to predict PV and UO was validated in a prohibitively small number of patients [20]–[25] or only at the population level [26]–[30]. In addition, the ability of these mathematical models to predict responses other than PV and UO are not reported in detail. Such lack of validity evidence weakens the credibility of these mathematical models as basis for reproducing PV and UO responses associated with wide-ranging burn patients. Third, a subset of the mathematical models (especially those reported early) [20]–[25] do not incorporate contemporary knowledge of burn-related physiology and pathophysiology, regarding in particular the burn-induced perturbations in blood volume kinetics and renal function, lymphatic flow, and tissue pressure-volume relationships [2], [31]–[33]. Closing these gaps may yield an enhanced mathematical model of burn injury and resuscitation better suited to the development and testing of novel burn resuscitation protocols and algorithms.

This paper presents a mathematical model of blood volume kinetics and renal function in response to burn injury and resuscitation, which is applicable to the development and non-clinical testing of burn resuscitation protocols and algorithms. Our mathematical model consists of an established multi-compartmental model of BV kinetics, a hybrid mechanistic-phenomenological model of renal function, and novel lumped-parameter models of burn-induced perturbations in volume kinetics and renal function equipped with contemporary knowledge on burn-related physiology and pathophysiology. We examined the efficacy of our mathematical model in predicting volume kinetic and renal function responses to burn injury and resuscitation using the dataset collected from 16 sheep.

This paper is organized as follows. Section Materials and Methods presents our mathematical model, experimental dataset, and data analysis details. Section Results presents results, which are discussed in Section Discussion. Section Conclusion concludes the paper with future directions.

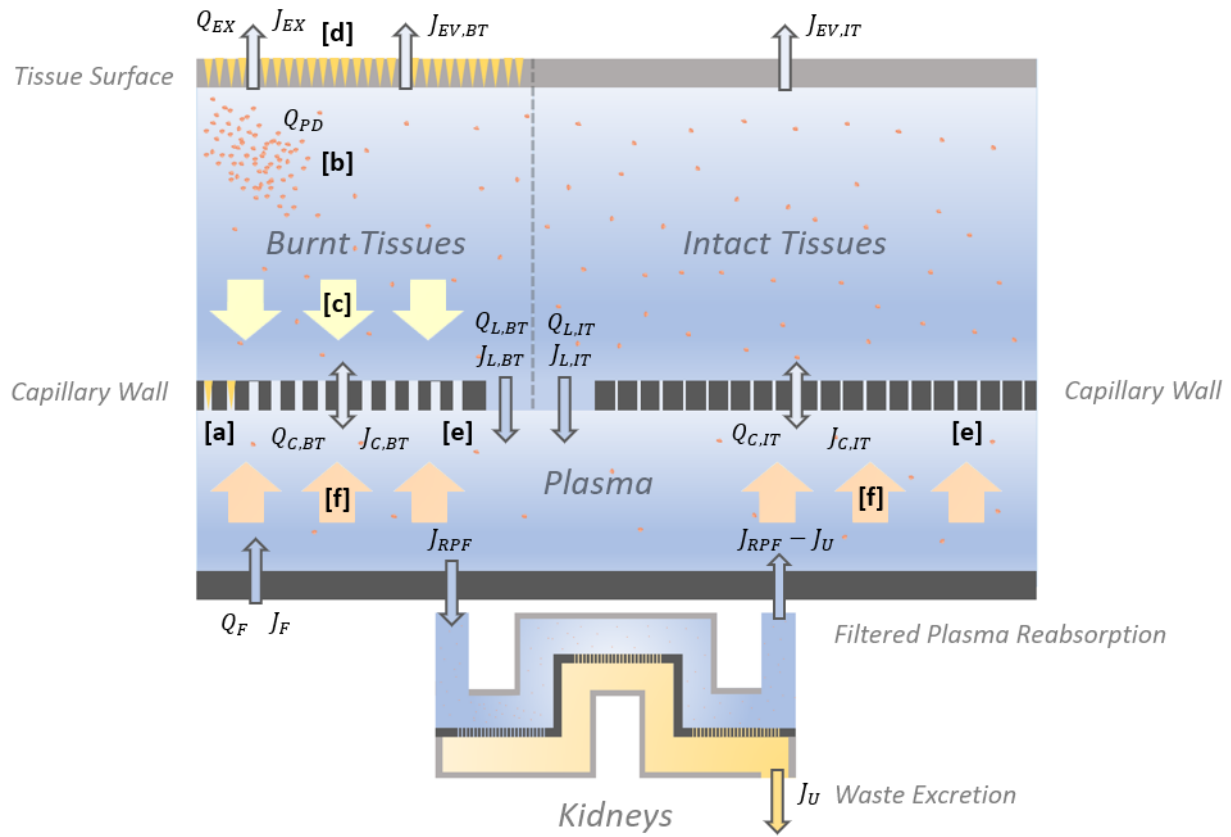

**Fig. 1**: Mathematical model to predict responses to burn injury and resuscitation. It includes volume kinetics to describe water volume and protein concentration in the vasculature and the tissues ("Plasma", "Burnt Tissues", "Intact Tissues", "Capillary wall", and "Tissue Surface"), renal function to describe UO response to vascular volume changes ("Kidneys"), and burn-induced perturbations in volume kinetics and renal function (denoted as "[a]" to "[f]"). J: water flow. Q: albumin flow. Subscripts: C (capillary filtration); L (lymph flow); F (fluid infusion); U (UO); RPF (renal plasma flow); EX (exudation); EV (evaporation); PD (protein denaturation); BT (burnt tissues); IT (intact tissues). [a]: Destruction of capillaries in burnt tissues. [b]: Denaturation of protein in burnt tissues. [c] Transient negative hydrostatic pressure in burnt tissues. [d]: Increased dermal fluid loss. [e]: Time-varying changes in capillary filtration and albumin transport. [f]: Vasodilation.

## MATERIAL AND METHODS

### Mathematical Model

Our mathematical model includes mechanisms and components to predict responses to burn injury and resuscitation: (i) volume kinetics to describe water volume and protein concentration in the vasculature and the tissues, (ii) renal function to describe UO response to vascular volume changes, and (iii) burn-induced perturbations in volume kinetics and renal function as a cascade of biochemical, molecular, and mechanical events (**Fig. 1**). Full details of the mathematical model, including all the equations, are provided in Appendix.

The volume kinetics (VK) was represented by a multi-compartmental model consisting of vasculature, intact tissues, and burnt tissues. It describes the fluid and protein balance in each compartment and homeostasis via capillary filtration and lymphatic flow as well as water gain (e.g., burn resuscitation) and loss (e.g., UO, evaporation, and exudation) (see Section A.1). The renal function was represented by a novel lumped-parameter model we developed in this work (**Fig. A1**; see Section A.2). It consists of hybrid mechanistic-phenomenological components to describe UO control by the kidneys, including the glomerular filtration rate (GFR) modulated by the Starling forces in response to the change in BV as well as the reabsorption and sodium osmosis modulated by the antidiuretic hormone (ADH). The burn-induced perturbations in VK and renal functions were represented by an array of phenomenological models, which describe local and systemic pathophysiological changes induced by burn injury in the form of time-varying perturbations acting on the parameters and variables in VK and renal function (shown as [a], [b], [c], [d],[e], and [f] in **Fig. 1**; see Section A.3).

## Experimental Data

Experimental dataset used to validate our mathematical model was formed by combining the two datasets collected from two prior work under the approval of local Institutional Animal Care and Use Committee [34], [35]. One dataset was obtained from a study conducted on adult sheep (N=8) with the median weight of 40kg and full-thickness burn injury of 40% total body surface area (TBSA). Burn resuscitation by lactated ringers was initiated 1 hour post burn and continued for 48 hours. Resuscitation was performed to maintain a target UO of 1-2ml/kg/h, which is considered as normal UO in sheep. Key measurements in the dataset used in this work include hourly records of fluid infusion and UO, and more sparse measurements of hematocrit (HCT; N=12 per sheep on the average). The other dataset was obtained from another study conducted on adult sheep (N=8) with the median weight of 50kg and full-thickness burn injury of 30% TBSA. Burn resuscitation by lactated ringers was initiated 2 hours post burn and continued for ≥24 hours, while sheep were monitored for 72 hours. Resuscitation was performed to restore and maintain central venous pressure and pulmonary wedge pressure. Key measurements in the dataset used in this work include fluid infusion, UO, and HCT. In a subset of the sheep, measurements of protein concentrations in plasma, burnt tissue, and intact tissue compartments, as well as fractions of lymphatic flow from burnt and intact skin to vascular compartment at lymph nodes were made. These measurements were also used to validate our mathematical model in this work. **Table 1** summarizes the measurements available in our dataset.

We computed PV from HCT using a formula reported in prior work [36], [37] based on the assumption that (i) the baseline BV of sheep is 63.5ml/kg and (ii) no blood loss occurred during the course of burn resuscitation (so that red blood cell volume was preserved).

**Table 1**: Measurement availability in dataset used for mathematical model validation. N: number of subjects associated with the measurement. PV: plasma volume. UO: urinary output. BT: burnt tissue. IT: intact tissue. Albumin: albumin concentration.

|   | Fluid Dose | PV | UO | Lymph Flow (BT) | Lymph Flow (IT) | Albumin (Plasma) | Albumin (BT) | Albumin (IT) |
|---|---|---|---|---|---|---|---|---|
| N | 16 | 16 | 16 | 6 | 8 | 5 | 2 | 2 |

## Data Analysis

We examined the validity of our mathematical model as follows. First, we categorized the parameters in the mathematical model into subject-invariant and subject-specific parameters. Second, we estimated the values of subject-specific parameters by fitting the mathematical model to the measurements in the dataset while fixing subject-invariant parameters to respective pre-specified values. Third, we examined the validity of the mathematical model by analyzing the goodness of fit and estimated parameter values associated with the mathematical model. Details follow.

First, we categorized the parameters in the mathematical model into subject-invariant and subject-specific parameters. Subject-invariant parameters included (i) those (mostly associated with extensive properties and mechanistic components in the mathematical model) whose values appear consistent in multiple prior literatures (e.g., nominal water volume and albumin content in the vascular and tissue compartments and the hydrostatic pressure in Bowman's capsule), and (ii) those whose values must be selected to produce mechanistically relevant physiological responses (e.g., parameters associated with the tissue compliance model, which must be chosen to yield physically relevant tissue hydrostatic pressure for a range of tissue volumes). The values of these subject-invariant parameters were mostly determined based on the existing literature (see **Table A1** for specific literatures we used to determine these values). Subject-specific parameters included (i) those whose values are anticipated to exhibit large inter-subject variability (e.g., burn-induced perturbations and nominal glomerular filtration coefficient), (ii) those whose values have rarely been reported in the existing literature (e.g., capillary elastance and nominal lymphatic flow), and (iii) those associated with phenomenological components in the mathematical model whose values are inherently unknown. After all, a total of 58 parameters were categorized into 34 subject-invariant and 24 subject-specific parameters.

Second, we estimated the values of subject-specific parameters by fitting the mathematical model to the measurements in the dataset while fixing subject-invariant parameters to respective pre-specified values. Considering that the amount of measurements in the dataset may not be sufficient to reliably estimate all the 24 subject-specific parameters, we selected a small subset of sensitive subject-specific parameters and estimated them while fixing the remaining insensitive subject-specific parameters to respective population-average values. First, we determined population-average parameter values by fitting our mathematical model simultaneously to all the measurements in all subjects in the dataset using the pooled approach [38]. Second, we estimated all 24 subject-specific parameters on the individual basis by fitting our mathematical model to the measurements associated with each subject, by employing a regularized fitting that intends to minimize the number of parametric deviations from the population-average values [39], [40]. Those subject-specific parameters exhibiting deviations from population-average values in many sheep were chosen as sensitive subject-specific parameters, which were confirmed via a post-hoc parametric sensitivity analysis. This exercise resulted in 11 sensitive subject-specific parameters that must be estimated on the individual basis (see Appendix A.5). Third, we estimated the 11 sensitive subject-specific parameters on the individual basis by fitting our mathematical model to all the available measurements associated with each subject while fixing the remaining 13 insensitive subject-specific parameters to respective population-average values (and as well, fixing subject-invariant parameters to respective pre-specified values). Details regarding the fitting of the mathematical model is provided in Appendix (see A.5). In sum, we derived 16 subject-specific mathematical models as well as a population-average mathematical model using the dataset.

Third, we examined the validity of the mathematical model by analyzing the goodness of fit and estimated parameter values associated with the mathematical model as follows. First, we examined the ability of our mathematical model to predict PV and UO responses on the individual basis, in terms of normalized mean absolute error (NMAE), correlation coefficient, and the Bland-Altman statistics between actual PV and UO measurements associated with each sheep in the dataset versus PV and UO predicted by our mathematical model characterized with the corresponding subject-specific parameter values. Second, we examined the ability of our mathematical model to predict physiologically plausible VK and renal function responses by (i) quantitatively analyzing, on the individual basis, NMAE, correlation coefficient, and the Bland-Altman statistics between actual lymphatic flow and albumin concentration associated with (a subset of) each sheep in the dataset (see **Table 1**) versus lymphatic flows and albumin concentrations predicted by our mathematical model characterized with the corresponding subject-specific parameter values, and also (ii) qualitatively comparing VK and renal function responses predicted by our mathematical model characterized by population-average parameters with contemporary knowledge on burn pathophysiology and findings from recent studies. Third, we examined the physiological plausibility of our mathematical model by comparing the subject-specific and population-average parameter values estimated by fitting the mathematical model to the dataset with known typical values and those reported in the literature.

## RESULTS

**Table 2** summarizes NMAE, correlation coefficient, and the Bland-Altman statistics (i.e., the limits of agreement) associated with PV, UO, lymphatic flow, and albumin concentration predicted by the mathematical model. **Fig. 2** presents representative examples of measured versus model-predicted PV and UO responses of two 40kg sheep subject to 40% burn whose PV was resuscitated (a) beyond the pre-burn level and (b) just to the pre-burn level. **Fig. 3** presents VK and renal function responses to burn injury and resuscitation predicted by population-average mathematical model in response to population-average burn resuscitation input.

**Table 2**: Normalized mean absolute error (NMAE; reported in median (IQR)), correlation coefficient (r), and Bland-Altman limits of agreement (LoA) associated with plasma volume, urinary output, lymphatic flow, and albumin concentration predicted by the mathematical model. PV: plasma volume. UO: urinary output. P: plasma. BT: burnt tissues. IT: intact tissues. LoA: 95% limits of agreement (bias+/-2×SD).

|  | PV [ml] (N=16) | UO [ml/h] (N=16) | Lymphatic Flow [ml/h] | | Albumin Concentration [g/l] | | |
|---|---|---|---|---|---|---|---|
|  |  |  | BT (N=6) | IT (N=7) | P (N=4) | BT (N=2) | IT (N=2) |
| NMAE [%] | 16 (3) | 17 (3) | 17 (7) | 17 (6) | 19 (10) | 7 (6) | 17 (11) |
| r | 0.82 | 0.66 | 0.92 | 0.91 | 0.85 | 0.98 | 0.85 |
| LoA | 50+/-404 | -8+/-70 | 0.2+/-5.8 | 4+/-56 | 0.5+/-8.2 | -0.07+/-3.7 | 2.3+/-4.0 |

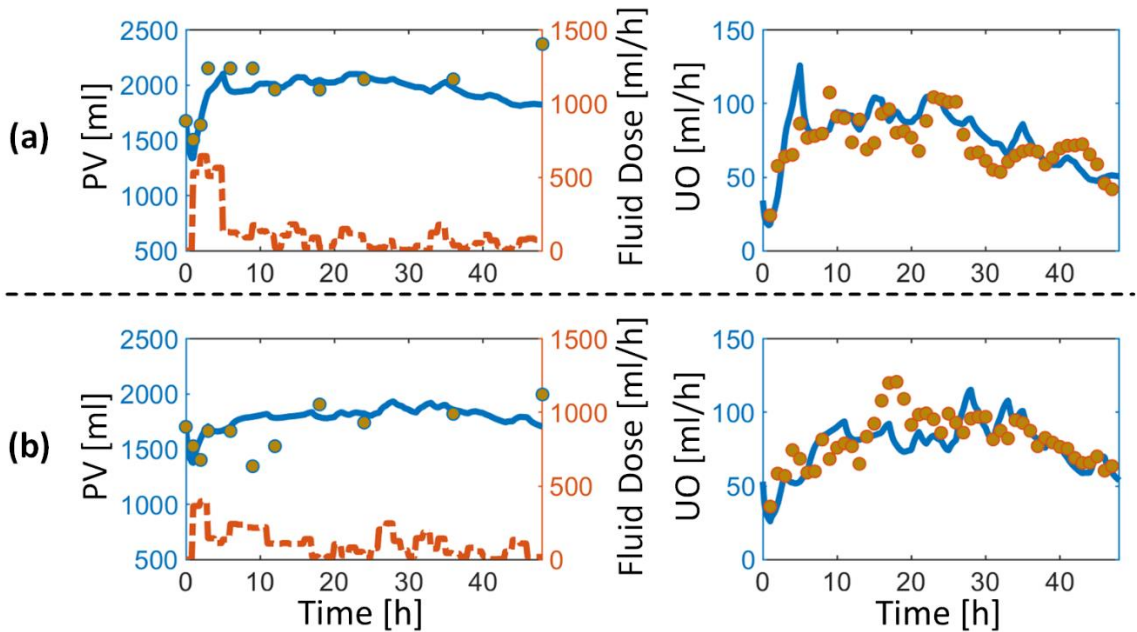

**Fig. 2**: Measured versus model-predicted plasma volume (PV) and urinary output (UO) responses of two 40kg sheep subject to 40% burn. Circles: measured responses. Solid lines: model-predicted responses. Dashed lines: measured fluid dose. (a) Sheep with PV resuscitated 200ml beyond pre-burn level 48 hours post-burn. (b) Sheep with PV resuscitated just up to pre-burn level 48 hours post burn.

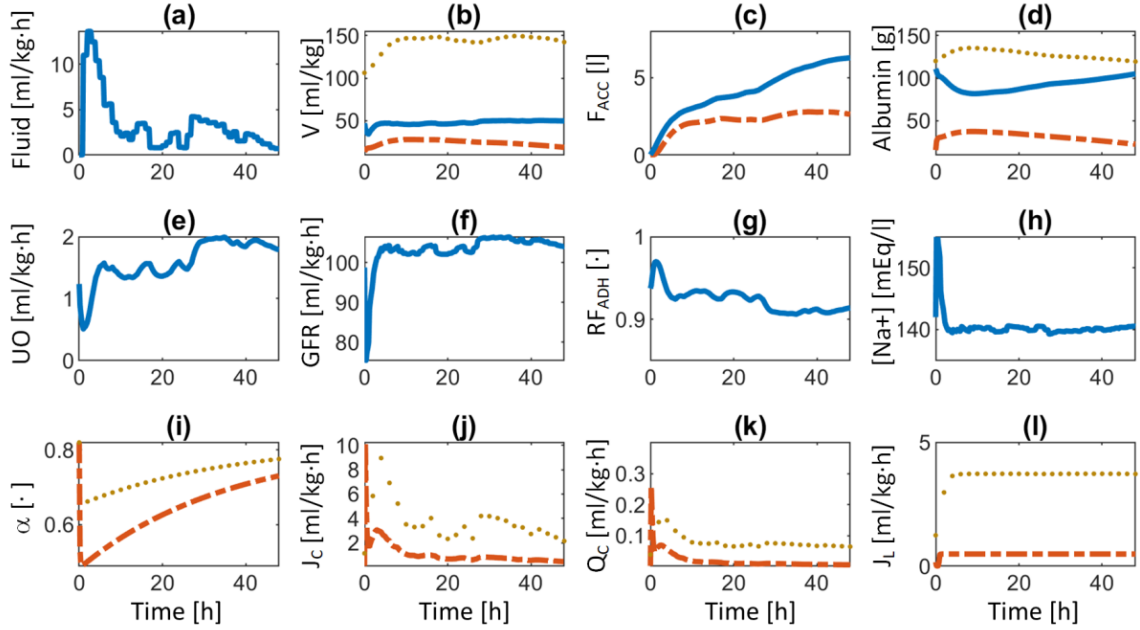

**Fig. 3**: Volume kinetic and renal function responses to burn injury and resuscitation predicted by population-average mathematical model. V: water volume. $F_{ACC}$: accumulated fluid. $RF_{ADH}$: reabsorption fraction due to ADH (see A.2). $\alpha$: capillary pore radius ratio (see A.3). $J_C$: capillary filtration. $Q_C$: albumin transport across the capillary wall. $J_L$: lymphatic flow. In (b) and (d), blue solid, orange dash-dot, and brown dotted lines are plasma volume as well as burnt and intact tissue volumes, respectively. In (c), blue solid and orange dash-dot lines are accumulated resuscitation fluid volume and fluid creep, respectively. In (i)-(l), orange dash-dot and brown dotted lines correspond to burnt and intact tissues, respectively.

## DISCUSSION

In the lack of consensus on the optimal burn resuscitation regimen and extreme inter-patient physiological variability in burn-induced responses, a credible mathematical model of burn injury and resuscitation may provide a meaningful basis for development and non-clinical testing of burn resuscitation protocols and algorithms in a wide range of patients. Existing mathematical models are associated with at least one of the following limitations: they are (i) not capable of predicting physiological responses essential in measuring the severity of burn injury and the effectiveness of resuscitation (including PV and UO), (ii) not rigorously validated to demonstrate the ability to capture formidable inter-patient variability in burn- and resuscitation-induced responses, and (iii) often equipped with obsolete knowledge of burn pathophysiology. Our goal was to develop a mathematical model of burn injury and resuscitation ideally suited to in-depth understanding of complex burn-induced VK and renal responses as well as development and non-clinical testing of burn resuscitation protocols and algorithms.

Our mathematical model could adequately predict PV and UO responses to burn injury and resuscitation (**Table 2** and **Fig. 2**). In particular, the NMAE associated with our UO prediction was considerably smaller than a recently reported black-box model [41] (30+/-6%) while the underlying UO variability was comparable (44% (our dataset) versus 38% [41] in terms of the coefficient of variation (CoV)). In addition, the level of prediction accuracy was reasonably consistent across subjects (3% in terms of IQR of NMAE for both PV and UO). Further scrutinizing UO prediction, there was 89% agreement between measured and predicted UO in terms of residing in the same range (<0.5ml/h/kg, 0.5-1.0ml/h/kg, and >1.0ml/h/kg) on the average. This is an encouraging performance given that contemporary burn resuscitation protocols adjust resuscitation dose based on UO range rather than its value. All in all, an important related implication is that our mathematical model may be able to capture inter-patient variability in burn-related pathophysiology. In fact, when characterized with individual-specific parameters, our mathematical model could reproduce largely distinct burn- and resuscitation-induced PV and UO responses of apparently similar sheep. For example, **Fig. 2** suggests that our mathematical model predicts PV and UO responses associated with two similar sheep of 40kg weight subject to the same 40% burn, but subject to largely distinct initial reduction in PV (approximately 300ml versus 500ml) and its subsequent recovery (200ml versus 1ml above initial PV level). Considering that VK (especially PV) and renal function (especially UO) are direct and surrogate measures of burn resuscitation, respectively, our mathematical model may be adequate for the intended context of use: development and testing of burn resuscitation protocols and algorithms.

In addition to PV and UO, our mathematical model could also predict various VK and renal responses to burn injury and resuscitation not readily accessible via routine clinical measurements in a physiologically plausible fashion. First, VK responses including protein concentrations and lymphatic flows predicted by our mathematical model exhibited adequate agreement with experimental measurements (**Table 2**). Second, all the VK and renal function responses predicted by our mathematical model exhibited behaviors qualitatively consistent with contemporary knowledge on burn pathophysiology and findings from recent studies in both individual and population-average senses. **Fig. 3** presents responses predicted by the population-average mathematical model. To mention a few, the model predicted that (i) PV and UO showed an anticipated trend of initial decline upon the onset the burn injury and subsequent recovery with resuscitation and also with the return of resuscitation fluid leaked into tissues back to vasculature >24 hours post burn (**Fig. 3(b)**, **Fig. 3(c)**, and **Fig. 3(e)**) [1], [42]; (ii) burnt tissue volume increased up to twice its initial value and peaked approximately at 10-20 hours post burn (**Fig. 3(b)**) [9]; (iii) intact tissue volume exhibited the same trend but with smaller degree (up to 30% above its initial value; **Fig. 3(b)**) [9]; (iv) plasma albumin was transported into burnt and intact tissues due to burn-induced perturbations in albumin reflection and permeability-surface area coefficients (**Fig. 3(d)** and **Fig. 3(k)**; see [e] in **Fig. 1**) as a result of burn-induced increase in the capillary pore size that decreased the capillary pore radius ratio in both burnt and intact tissue (**Fig. 3(i)**) [43]; and (v) GFR increased just hours post burn even before PV was restored [44] (**Fig. 3(f)**). All in all, its ability to make physiologically plausible predictions of VK and renal function makes our mathematical model suited for gaining sophisticated insights not directly available in routine clinical measurements (e.g., UO). For example, our mathematical model predicts that on the average fluid creep peaks at approximately 10 hours post burn and that more than 50% of the resuscitation fluid in the first 24 hours leaks into tissues to exacerbate edema instead of contributing to hemodynamic recovery, which is in good agreement with the findings in the literature (**Fig. 3(b)** and **Fig. 3(c)**) [1], [14]. It also predicted that sodium concentration decreased after burn injury and resuscitation, which is also consistent with the findings in the literature [45] (**Fig. 3(h)**). Provided that rigorous development and testing of burn resuscitation protocols and algorithms require comprehensive understanding and in-depth scrutiny of complex VK and renal function responses to burn injury and resuscitation, this attribute may confer additional credibility for the intended context of use on our mathematical model.

It is worth noting that our mathematical model could predict adequate PV and UO as well as physiologically plausible VK and renal function responses while characterized with physiologically acceptable parameter values. In fact, the majority of model parameters associated with physiological implications assumed values comparable to typical values and/or those reported in the literature both on the individual and population-average basis (**Table 3**). Such a physiological acceptability in the model parameter values suggests physiological plausibility (or perhaps even relevance) of our mathematical model, especially the novel mechanisms and components we developed (e.g., our representations for the renal function and burn-induced VK and renal perturbations). It is also worth noting that our mathematical model embraces a wealth of contemporary knowledge on burn physiology and pathophysiology based on findings in recent literatures, including but not limited to burn-induced systemic increase in capillary pore radius [32] and GFR [44] as well as highly nonlinear tissue compliance characteristics dictated by the glycosaminoglycans (GAG) properties [46]. In fact, many of these new findings were made after the pioneering work by Arturson et al. [20]–[22], Roa et al. [23]–[25], and Bert et al. [26]–[30] was conducted. Hence, they were inevitably not incorporated in the existing mathematical models developed by these pioneering researchers. The physiological and mechanistic plausibility of our mathematical model combined with its incorporation of a range of contemporary knowledge on burn pathophysiology allows our mathematical model to predict PV and UO as well as many VK and renal function responses essential to develop and test burn resuscitation protocols and algorithms against a wide range of pathophysiological variability.

Our work has a number of limitations. First, we (somewhat implicitly) assumed that fluid transfers between VK compartments are isotonic and that VK compartments are in electrolyte balance at all times. These are tenable assumptions in that (i) electrolytes pass capillary walls easily, (ii) they are mixed with water quickly and continuously, and (iii) burn resuscitation fluid considered in this work is isotonic lactated ringers. Hence, we assumed that the primary factor governing the changes in electrolyte concentration in the body is related to the change in sodium concentration due to water reabsorption in collecting ducts in the kidneys (which was considered in our renal function model; see Section A.2). Our mathematical model was able to predict VK and renal function responses reasonably well despite these simplifying assumptions, which suggests that the assumptions may be adequate (or at least may not have had drastic impact on the efficacy of the mathematical model). However, these assumptions may not be valid in case of burn resuscitation with fluid other than lactated ringers. Hence, future work on expanding our mathematical model by removing these assumptions is required. Second, our mathematical model only includes extracellular compartments but not intracellular compartment. As a matter of fact, we initially considered intracellular compartment due to its potential importance to electrolyte balance, but it ended up with adding unnecessary complexity to the mathematical model without any meaningful improvement in the goodness of fit. Regardless, intracellular compartment may need to be incorporated in order to broaden the applicability of our mathematical model beyond lactated ringers. Hence, future work on expanding our mathematical model to include intracellular compartment model of adequate complexity and efficacy may be rewarding. Third, the dataset used in this work was associated with rather uniform injury severity (30%-40% TBSA). Despite a wide range of VK and renal function responses in the dataset (Table 1), the narrow range of injury severity may have prevented us from garnering additional insight into, e.g., the dependence of burn-induced perturbations on the injury severity. Hence, future work is required to investigate the adequacy of our mathematical model (especially its phenomenological models of burn-induced VK and renal function perturbations) under a wide range of burn injury severity, and if needed, to improve the validity and efficacy of our mathematical model against wide-ranging burn injury severity. Fourth, the mathematical model was not validated using extensive and ideal VK and renal function measurements. In particular, PV was derived from hematocrit measurements. Despite its well-known direct relationship to PV, its accuracy is not always perfect and is also impacted by disturbances such as hemolysis. In addition, UO was the only renal function measurement used to validate the mathematical model. We illustrated that our mathematical model, by virtue of mechanistic components therein, can at least predict a large number of VK and renal function variables that are qualitatively adequate (**Fig. 3**). Yet, future work must scrutinize the validity of our mathematical model using robust measurements of more extensive set of VK and renal function measurements.

## CONCLUSIONS

We developed a mathematical model of burn injury and resuscitation intended for use in the development and non-clinical testing of burn resuscitation protocols and algorithms. Using the dataset associated with sheep, we demonstrated the potential of our mathematical model for such context of use: it could predict PV and UO as well as a range of VK and renal function responses to burn injury and resuscitation by virtue of its physiological and mechanistic relevance combined with contemporary knowledge of burn physiology and pathophysiology. In order to establish its efficacy as a powerful non-clinical tool for developing and testing burn resuscitation protocols and algorithms, effort must be invested to validate our mathematical model in real-world burn patients associated with a wide range of physiological variability, injury severity, and resuscitation protocol.

## APPENDIX: MATHEMATICAL MODEL DETAILS

### A.1. Volume Kinetics

The VK was represented by a classical multi-compartmental model consisting of vasculature, intact tissues, and burnt tissues. It describes the water and protein balance in these compartments and homeostasis via capillary filtration and lymphatic flow as well as water gain (e.g., burn resuscitation) and loss (e.g., UO, evaporation, and exudation). The water and protein balance in each compartment was formulated based on the mass conservation principle:

$$\frac{d(V_P)}{dt} = -J_{C,BT} - J_{C,IT} + J_{L,BT} + J_{L,IT} + J_F - J_U, \tag{1a}$$

$$\frac{d(V_{BT})}{dt} = J_{C,BT} - J_{L,BT} - J_{EX} - J_{EV,BT}, \tag{1b}$$

$$\frac{d(V_{IT})}{dt} = J_{C,IT} - J_{L,IT} - J_{EV,IT}, \tag{1c}$$

$$\frac{d(A_P)}{dt} = -Q_{C,BT} - Q_{C,IT} + Q_{L,BT} + Q_{L,IT} + Q_F, \tag{2a}$$

$$\frac{d(A_{BT})}{dt} = Q_{C,BT} - Q_{L,BT} - Q_{EX} + Q_{PD}, \tag{2b}$$

$$\frac{d(A_{IT})}{dt} = Q_{C,IT} - Q_{L,IT}, \tag{2c}$$

where $V$ is water volume, $A$ is albumin content, $J$ is water flow, and $Q$ is albumin flow. The subscripts $C$, $L$, $F$, $U$, $EV$, $EX$, and $PD$ denote capillary filtration, lymphatic flow, fluid infusion, UO, evaporation, exudation, and protein influx due to burn-induced denaturation (see A.3), respectively, while the subscripts $P$, $BT$, $IT$ denote plasma, burnt tissues, and intact tissues, respectively. Note that we assumed that albumin content represents the protein content similarly to prior work [30], [47].

The capillary filtration was expressed using the Starling equation:

$$J_{C,X} = K_{C,X}[P_C - P_X - \sigma_X(\pi_C - \pi_X)], \tag{3}$$

where $X \in \{BT, IT\}$, $P_C$ and $P_X$ are the capillary and tissue hydrostatic pressures, $\pi_C$ and $\pi_X$ are the plasma and tissue colloid oncotic pressures, and $K_{C,X}$ and $\sigma_X$ are the capillary filtration coefficient and the albumin reflection coefficient associated with burnt ($X = BT$) and intact ($X = IT$) tissues, respectively. The colloid oncotic pressure was expressed as a linear function of albumin concentration [47]:

$$\pi_C = C_O[A]_P, \ \pi_X = C_O[A]_X, \tag{4}$$

where $C_O$ is a constant relating albumin concentration to colloid oncotic pressure, and $[A]_P = \frac{A_P}{V_P}$ and $[A]_X = \frac{A_X}{V_X}$ are the plasma and tissue albumin concentrations, respectively. The lymphatic flow was expressed by a phenomenological model in the form of a sigmoidal curve:

$$J_{L,X} = \frac{\bar{J}_{L,X}}{C_L + (1 - C_L)e^{-S_L(P_X - \bar{P}_X)}}, \tag{5}$$

where $\bar{J}_{L,X}$ is the nominal lymphatic flow at the nominal tissue hydrostatic pressure of $\bar{P}_X$, and $C_L$ and $S_L$ are constants representing the maximal degree of increase in the lymphatic flow and the its sensitivity to the change in the tissue hydrostatic pressure. Note that this phenomenological model can reproduce the real-world behavior of the lymphatic flow despite its simplicity: (i) it primarily depends on the tissue hydrostatic pressure [48]; (ii) it is proportional to the tissue hydrostatic pressure but eventually saturates; and (iii) it reduces to zero at very low tissue hydrostatic pressure [49].

The albumin transport across the capillary wall was expressed based on the coupled diffusion-convection equation [47], [50]:

$$Q_{C,X} = J_{C,X}(1 - \sigma_X) \left\{ \frac{[A]_P - [A]_X e^{-\frac{(1-\sigma_X)J_{C,X}}{PS_X}}}{1 - e^{-\frac{(1-\sigma_X)J_{C,X}}{PS_X}}} \right\}, \qquad (6)$$

where $X \in \{BT, IT\}$, and $PS$ is the permeability-surface area coefficient. The albumin transport associated with the lymphatic flow was expressed in terms of the lymphatic flow and the albumin concentration in the tissues:

$$Q_{L,X} = J_{L,X}[A]_X. \qquad (7)$$

The evaporation flow ($J_{EV,X}$, $X \in \{BT, IT\}$ in Eq. (1b)-(1c)) and the exudation flow ($J_{EX}$ in Eq. (1b) and $Q_{EX}$ in Eq. (2b)) were expressed as phenomenological models based on prior work (see Section A.3). The burn resuscitation ($J_F$ in Eq. (1a) and $Q_F$ in Eq. (2a)) is the input provided to the mathematical model.

The capillary ($P_C$) and tissue ($P_X$) hydrostatic pressures in Eq. (3)-(5) were expressed as functions of the corresponding water volumes ($V_P$ and $V_X$). We assumed a linear phenomenological relationship between $P_C$ and $V_P$:

$$P_C = \bar{P}_C + E_C(V_P - \bar{V}_P), \qquad (8)$$

where $\bar{P}_C$ is the nominal capillary hydrostatic pressure associated with the nominal PV $\bar{V}_P$, and $E_C$ is the capillary elastance. We used a nonlinear mechanistic hydrostatic pressure-volume relationship associated with the interstitial tissues at the microscopic level developed by Øien and Wiig [46], and extended it for use with macroscopic measurements. The original model by Øien and Wiig expresses the tissue hydrostatic pressure as a function of the radius $R$ of the spherical glycosaminoglycans (GAG's) therein:

$$P_X = -\frac{\alpha}{R} + \gamma, \qquad (9)$$

where $\alpha$ and $\gamma$ are constant coefficients representing the impact of electrostatic pressure and tissue tension pressure [46] on $P_X$ and

$$R = R(y_X) = \hat{R}\left[1 - (1 - \beta)(\frac{y_X - \hat{y}}{\check{y} - \hat{y}})\right]^n, \qquad (10)$$

where $y_X$ is the half-thickness of the extracellular matrix building block (which is a measure of hydration in the tissue $X$ at the microscopic level), $\hat{y}$ and $\check{y}$ are the maximum and minimum values of $y_X$, $\hat{R}$ is the maximum value of $R$, $\beta$ is the ratio between maximum $R$ and nominal $R$, and $n$ is an exponent describing the GAG response to hydration [46]. We extended Eq. (9)-(10) to compute $P_X$ from macroscopic rather than microscopic (i.e., $y_X$) measurement by assuming that $y_X$ is proportional to the water volume in the tissues:

$$y_X = 0.75\left(1 + \bar{W}_X \frac{V_X}{\bar{V}_X}\right), \qquad (11)$$

where $\bar{W}_X$ is the nominal hydration level defined as the ratio of the water volume in $X$ and its dry weight [33], and the coefficient 0.75 was derived from the assumptions used in prior work of Øien and Wiig [33].

## A.2. Renal Function

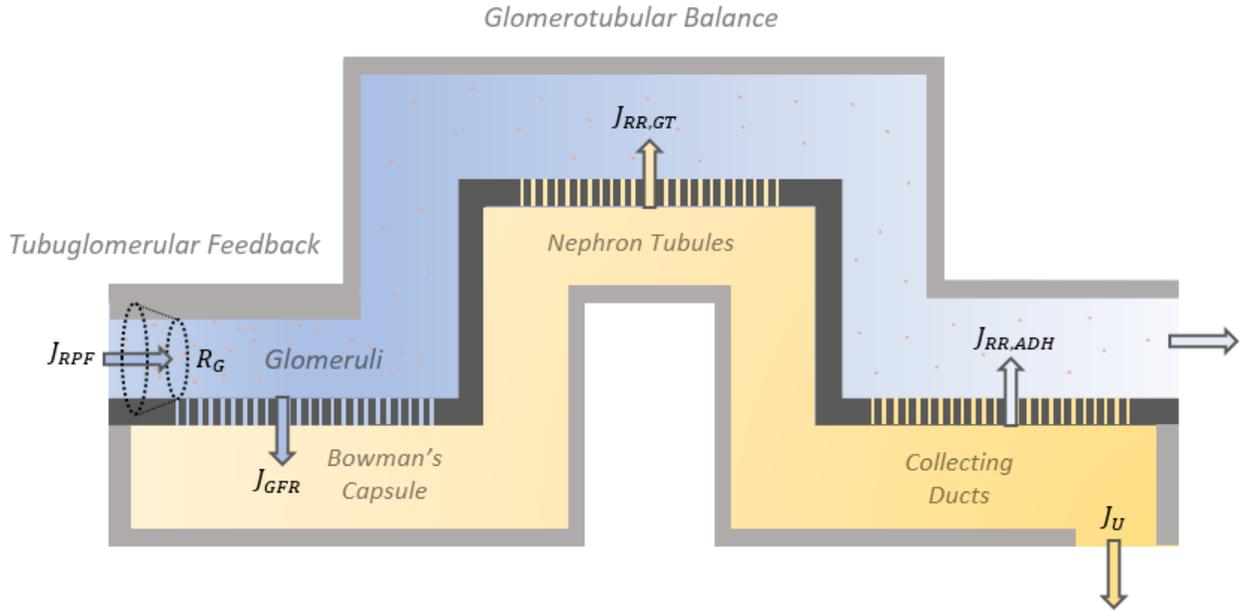

**Fig. A1**: Schematic of renal function model. $J_{RPF}$: renal plasma flow. $J_{GFR}$: glomerular filtration rate (GFR). $J_{RR,GT}$: reabsorption rate due to modulation of glomerulotubular balance. $J_{RR,ADH}$: reabsorption rate due to modulation of ADH. $J_U$: UO. $R_G$: hypothetical glomerular resistance modulating $J_{RPF}$ against perturbations in plasma volume.

The renal function was represented by a novel lumped-parameter, hybrid mechanistic-phenomenological model we developed in this work (**Fig. A1**). It consists of hybrid mechanistic-phenomenological components to describe UO control by the kidneys, including the glomerular filtration rate (GFR) modulated by the Starling forces in response to the change in BV as well as the reabsorption and sodium osmosis modulated by the antidiuretic hormone (ADH). UO is given by the difference between the GFR ($J_{GFR}$) and the reabsorption rate (RR; $J_{RR}$):

$$J_U = J_{GFR} - J_{RR}. \tag{12}$$

### A.2.1. Glomerular Filtration
GFR is dictated by the Starling forces:

$$J_{GFR} = K_G[P_G - P_B - \sigma_G(\pi_G - \pi_B)] \approx K_G[P_G - P_B - \pi_G], \tag{13}$$

where $P_G$ and $P_B$ are the hydrostatic pressures associated with the glomerular capillaries and the Bowman's capsules, $\pi_G$ and $\pi_B$ are the colloid oncotic pressures associated with the glomerular capillaries and the Bowman's capsules, $K_G$ is the glomerular filtration coefficient, and $\sigma_G$ is the glomerular albumin reflection coefficient. Noting that the membranes in the glomerular capillaries do not allow the passage of albumin, we assumed $\sigma_G = 1$ and $\pi_B = 0$ in Eq. (13).

$P_G$ depends on the renal plasma flow (RPF), a fraction of CO, which is perturbed by the fluctuations in PV but is also regulated by the tubulo-glomerular feedback (TGF) [51]. Assuming that CO is proportional to PV:

$$J_{RPF} = \bar{J}_{RPF} \frac{V_P}{\bar{V}_P}, \tag{14}$$

where $J_{RPF}$ is RPF, and $\bar{J}_{RPF}$ is its nominal value corresponding to the nominal PV $\bar{V}_P$. The perturbations in $J_{RPF}$ due to the fluctuations in PV induced by burn injury and resuscitation are strictly compensated for with the modulation of the renal capillary resistance (called the glomerular resistance [52]) controlled by TGF. We expressed TGF as simple phenomenological dynamic modulation of a hypothetical glomerular resistance $R_G$:

$$\tau_{TGF} \frac{d\Delta R_G}{dt} = -\Delta R_G + \frac{K_{TGF}}{\bar{J}_{GFR}} (J_{GFR} - \bar{J}_{GFR}), \tag{15}$$

where $\Delta R_G = R_G - \bar{R}_G$ is the modulation of $R_G$ from its nominal value $\bar{R}_G$ enforced by TGF, $\tau_{TGF}$ and $K_{TGF}$ are the time constant and the sensitivity associated with TGF, and $\bar{J}_{GFR}$ is the nominal value of GFR. Noting that $P_G$ is proportional to RPF and inversely proportional to $R_G$ (since an increase in $R_G$ results in a decrease in RPF), we expressed $P_G$ as a simple phenomenological function of $J_{RPF}$ and $R_G$:

$$P_G = \lambda_G \frac{J_{RPF}}{R_G}, \tag{16}$$

where $\lambda_G$ is a constant coefficient representing the sensitivity of $P_G$ to $J_{RPF}$ and $R_G$.

We assumed that $P_B$ is constant. Indeed, $P_B$ hardly varies except in rare cases, e.g., when the urinary tract is obstructed [53].

We expressed $\pi_G$ as a function of $\pi_C$ (see Eq. (3)) based on the idealistic assumption that albumin content is conserved in the renal capillaries and in the glomerular capillaries:

$$[A]_C J_{RPF} = [A]_G (J_{RPF} - J_{GFR}) \rightarrow [A]_G = \frac{J_{RPF}}{J_{RPF} - J_{GFR}} [A]_C = \frac{1}{1 - \varepsilon_{GFR}} [A]_C, \tag{17}$$

where $[A]_G$ is the glomerular albumin concentration and $\varepsilon_{GFR} = \frac{J_{GFR}}{J_{RPF}}$ is the filtration fraction [54]. Hence, based on the assumption that $C_{O,G} = C_O$, i.e., the relationship between albumin concentration and colloid oncotic pressure is the same in the renal capillaries and in the glomerular capillaries, $\pi_G$ can be expressed by the following for small $\varepsilon_{GFR}$:

$$\pi_G = \frac{1}{1 - \varepsilon_{GFR}} \pi_C \approx (1 + \varepsilon_{GFR}) \pi_C. \tag{18}$$

Plugging Eq. (16) and Eq. (18) into Eq. (13) yields the following expression for GFR:

$$J_{GFR} \approx \frac{K_G J_{RPF}}{J_{RPF} + K_G \pi_C} \left[ \lambda \frac{J_{RPF}}{R_G} - P_B - \pi_C \right]. \tag{19}$$

Note that GFR is now expressed in terms of PV ($J_{RPF}$; see Eq. (14)) and plasma albumin content ($\pi_C$; see Eq. (4)). Hence, our hybrid mechanistic-phenomenological model of GFR allows us to represent GFR using macroscopic VK.

### A.2.2. Reabsorption
RR ($J_{RR}$) is dictated by the modulation of glomerulotubular balance (GT; $J_{RR,GT}$) and ADH ($J_{RR,ADH}$):

$$J_{RR} = J_{RR,GT} + J_{RR,ADH}. \tag{20}$$

GT modulates the proximal tubules so that approximately 20% of GFR reaches the collecting ducts ($J_{RR,GT} \approx 0.8 J_{GFR}$) [55], [56]. The ADH content is modulated by baroreceptor (due to changes in BV) and osmoreceptor (due to changes in the sodium concentration ([Na$^+$]) in the blood) to control RR at the collecting ducts [54], [57]. We expressed the dynamics of the ADH content by the following phenomenological model:

$$\frac{d(ADH)}{dt} = K_{ADH} e^{\left(-\lambda_{V_P} \Delta V_P + \lambda_{[Na^+]} \Delta [Na^+]\right)} - 0.27 K_{ADH} [ADH] \frac{J_{GFR}}{\bar{J}_{GFR}} - 0.73 K_{ADH} [ADH] \frac{V_P}{\bar{V}_P}, \quad (21)$$

where $ADH$ is the ADH content in the extracellular fluid, $K_{ADH}$ is the nominal ADH secretion rate, $\lambda_{V_P}$ and $\lambda_{[Na^+]}$ are positive constant coefficients representing the sensitivity of ADH secretion on the changes in PV and [Na$^+$], and $[ADH]$ is ADH concentration in the extracellular fluid:

$$[ADH] = \frac{ADH}{V_P + V_{BT} + V_{IT}}. \quad (22)$$

Eq. (20) indicates that ADH secretion increases if PV decreases and/or [Na$^+$] increases [54]. The coefficients 0.27 and 0.73 come from the notion that ADH is excreted in the kidneys (27%; proportional to GFR [58]) and the liver (73%; proportional to hepatic blood flow, and therefore, approximately to PV [59]). ADH modulates the reabsorption fraction (RF) $\varepsilon_{RR} = \frac{J_{RR,ADH}}{J_{GFR} - J_{RR,GT}}$ to control RR at the collecting ducts. We adopted the Michaelis-Menten equation to express the relationship between $[ADH]$ and the $\varepsilon_{RR}$ [57]:

$$\varepsilon_{RR} = K_{RR} \frac{[ADH]}{[ADH]_{50} + [ADH]}, \quad (23)$$

where $K_{RR}$ is the maximum RF, and $[ADH]_{50}$ is the ADH concentration corresponding to $\varepsilon_{RR} = \frac{1}{2} K_{RR}$.

We computed $[Na^+]$ in Eq. (21) based on the idealistic assumption that sodium content is conserved in the body (including the collecting ducts) after burn injury and resuscitation:

$$[Na^+] = \frac{\bar{J}_{RR,ADH}}{J_{RR,ADH}} \overline{[Na^+]}, \quad (24)$$

where $\overline{[Na^+]}$ is the nominal sodium concentration corresponding to the nominal RR $\bar{J}_{RR,ADH}$. That the change in $[Na^+]$ primarily depends on $J_{RR,ADH}$ may be justified to an extent because $J_{RR,ADH}$ consists of pure water diluting the plasma [57].

Substituting Eq. (19), Eq. (20), and Eq. (23) along with $J_{RR,GT} \approx 0.8 J_{GFR}$ into Eq. (12) yields the following expression for UO:

$$J_U = J_{GFR} - J_{RR} = J_{GFR} - J_{RR,GT} - J_{RR,ADH} \approx 0.2 \frac{K_G J_{RPF}(V_P)}{J_{RPF} + K_G \pi_C} \left[ \lambda \frac{J_{RPF}(V_P)}{R_G} - P_B - \pi_C \right] (1 - \varepsilon_{RR}). \quad (25)$$

Note that UO is now expressed in terms of PV ($J_{RPF}$; see Eq. (14)) and plasma albumin content ($\pi_C$; see Eq. (4)), and ADH concentration. Hence, our hybrid mechanistic-phenomenological model of UO allows us to represent UO using macroscopic VK and ADH dynamics.

## A.3. Burn-Induced Pathophysiology in Volume Kinetics and Renal Function

The burn-induced pathophysiological perturbations in VK and renal functions were represented by an array of phenomenological models, which describe local and systemic pathophysiological changes induced by burn injury in the form of time-varying perturbations acting on the parameters and variables associated with VK and renal function. Local perturbations, restricted to burnt tissues, include: (i) destruction of capillaries in burnt tissues [2], (ii) denaturation of protein in burnt tissues [31], [43], [60], (iii) transient negative hydrostatic pressure in burnt tissues [60], [61], and (iv) dermal fluid loss [62], [63]. Systemic perturbations include: (i) time-varying changes in capillary filtration and albumin transport [32] and (ii) vasodilation [64].

We devised a universal function, $\phi(t)$, to represent the time-varying perturbations in all the VK and renal function parameters and variables:

$$\phi(M_W, \lambda_{1,W}, \lambda_{2,W}, t) = M_W(e^{-\lambda_{1,W}t} - e^{-\lambda_{2,W}t}), \tag{26}$$

where $M_W$ is the maximum perturbation (occurring after the onset of burn injury), $\lambda_{1,W}$ and $\lambda_{2,W}$ are the slow time constant and fast time constant associated with the decay of the perturbation, all corresponding to a specific perturbation $W$. The parameter $\mu$ denotes the ratio of $\lambda_{2,W}$ to $\lambda_{1,W}$. We used Eq. (26) to express systemic perturbations as well as transient negative hydrostatic pressure in burnt tissues: $W \in \{\alpha_{BT}, \alpha_{IT}, P_C, P_{BT}\}$, where $\alpha_{BT}$ and $\alpha_{IT}$ are pore ratios associated with burnt and intact tissues (see below for details).

### A.3.1. Local Pathophysiology

We expressed the destruction of capillaries in burnt tissues as a decrease in the capillary filtration coefficient $K_{C,BT}$ and permeability surface area coefficient $PS_{BT}$ by a factor of $k_{PD\_X}$ (see Eq. (32)) [2]. We expressed the denaturation of protein as a protein influx $Q_{PD}$ into burnt tissues [31], [43], [60] (see Eq. 2b):

$$Q_{PD} = \hat{Q}_{PD} e^{-\lambda_{PD} t}, \tag{27}$$

where $\hat{Q}_{PD}$ is the protein influx immediately post burn, decaying with a time constant of $\lambda_{PD}$. We expressed the transient negative hydrostatic pressure in burnt tissues using $\phi(t)$ in Eq. (26) as follows:

$$\Delta P_{BT}(t) = -\phi(M_{P_{BT}}, \lambda_{1,P_{BT}}, \lambda_{2,P_{BT}}, t), \tag{28}$$

where $\Delta P_{BT}(t)$ is the burn-induced perturbation in the hydrostatic pressure in burnt tissues. The overall hydrostatic pressure in burnt tissues is computed by combining Eq. (9) and Eq. (28):

$$P_{BT}(t) = -\frac{\alpha}{R(y_{BT})} + \gamma + \Delta P_{BT} = -\frac{\alpha}{R(y_{BT})} + \gamma - \phi(M_{P_{BT}}, \lambda_{1,P_{BT}}, \lambda_{2,P_{BT}}, t). \tag{29}$$

We used phenomenological models of evaporation and exudation reported in prior work [65], [66]:

$$J_{EV,BT} = \begin{cases} K_{1,EV} \varepsilon_B S_B e^{\lambda_{1,EV} t}, & t < 6\ hr \\ K_{2,EV} \varepsilon_B S_B e^{\lambda_{2,EV} t}, & t > 6\ hr \end{cases}, \tag{30a}$$

$$J_{EV,IT} = K_{1,EV}(1 - \varepsilon_B) S_B, \tag{30b}$$

$$J_{EX} = K_{EX} \varepsilon_B S_B e^{\lambda_{EX} t}, \tag{31a}$$

$$Q_{EX} = J_{EX} \eta_{EX} [A]_{BT}, \tag{31b}$$

where $K_{1,EV} > 0$, $K_{2,EV} > 0$ and $K_{EX} > 0$ are constant coefficients, and $\lambda_{2,EV} > 0$, $\lambda_{2,EV} < 0$ and $\lambda_{EX} < 0$ are time constants, $\varepsilon_B$ is the fraction of body surface subject to burn, $S_B$ is TBSA, and $\eta_{EX}$ is the ratio between the albumin concentration in the exudate and the albumin concentration in the burnt tissues [30].

### A.3.2. Systemic Pathophysiology

We expressed the systemic impact of burn injury as time-varying changes in the capillary filtration, albumin reflection, and permeability-surface area coefficients (representing perturbations in capillary filtration and albumin transport) as well as capillary hydrostatic pressure (representing perturbation in vasodilation). By capitalizing on the pore theory of trans-capillary exchange [67], [68], we expressed the capillary filtration, albumin reflection, and permeability-surface area coefficients associated with burnt and intact tissues as functions of the pore radius ratios $\alpha_X$, $X \in \{BT, IT\}$, defined as the ratio between the albumin radius and capillary pore radius:

$$K_{C,X} = \overline{K}_{C,X} k_{PD,X} \frac{\bar{\alpha}_X^4}{\alpha_X^4}, \tag{32a}$$

$$\sigma_X = 1 - (1 - \alpha_X)^2, \tag{32b}$$

$$PS_X = \overline{PS}_X k_{PD,X} \frac{\bar{\alpha}_X^2 (1 - \alpha_X^2)}{(1 - \bar{\alpha}_X^2) \alpha_X^2}, \tag{32c}$$

where $\overline{K}_{C,X}$, $\overline{PS}_X$, and $\bar{\alpha}_X$ are the nominal values of $K_{C,X}$, $PS_X$, and $\alpha_X$, adjusted for the water fraction in burnt and intact tissues, weight, and capillary recruitment [30]:

$$\overline{K}_{C,BT} = \overline{K}_C \varepsilon_B r_{FV} \eta_{CR}, \ \overline{K}_{C,IT} = \overline{K}_C (1 - \varepsilon_B r_{FV}) \eta_{CR}, \tag{33a}$$

$$\overline{PS}_{BT} = \overline{PS} \varepsilon_B r_{FV} \eta_{CR}, \ \overline{PS}_{IT} = \overline{PS} (1 - \varepsilon_B r_{FV}) \eta_{CR}, \tag{33b}$$

where $\overline{K}_C$ and $\overline{PS}$ are nominal capillary filtration and permeability surface area coefficients in the absence of burn injury, $\eta_{CR} = \left(2 \frac{V_P}{\overline{V}_P} - 1\right)$, and $r_{FV}$ is the fluid volume ratio between the skin and the total interstitial compartment [30].

Note that $k_{PD} = 1$ if $X = IT$, because capillary destruction does not occur in intact tissues. We expressed the burn-induced changes in these coefficients by formalizing the burn-induced changes in the capillary pore radius ratios using Eq. (26):

$$\alpha_X(t) = \bar{\alpha}_X - \phi(M_{\alpha_X}, \lambda_{1,\alpha_X}, \lambda_{2,\alpha_X}, t). \tag{34}$$

We likewise expressed the burn-induced change in capillary hydrostatic pressure [64] using Eq. (26):

$$\Delta P_C(t) = \phi(M_{P_C}, \lambda_{1,P_C}, \lambda_{2,P_C}, t). \tag{35}$$

The overall capillary hydrostatic pressure is computed by combining Eq. (8) and Eq. (35):

$$P_C = \bar{P}_C + E_C(V_P - \bar{V}_P) + \Delta P_C(t) = \bar{P}_C + E_C(V_P - \bar{V}_P) + \phi(M_{P_C}, \lambda_{1,P_C}, \lambda_{2,P_C}, t). \tag{36}$$

## A.4. Model Parameters: Nomenclature, Definitions and Values

**Table A1**: Mathematical model parameters: definitions, categories (I/S), and values. I: subject-invariant parameters. S: subject-specific parameters. The vales are given as mean, median (IQR), or mean+/-SD.

| Symbol | Definition | I/S | Value (Model) | Value (Literature) |
|---|---|---|---|---|
| $\overline{BV}$ | Nominal blood volume [ml/kg] | I | 63.5 | 63.5 [69] |
| $\bar{V}_P$ | Nominal water volume in plasma [ml/kg] | I | 42.8 | 42 [70]-46 [30] |
| $r_{FV}$ | Skin fluid volume to total interstitial fluid volume ratio [·] | I | 0.28 | 0.28 [30] |
| $\bar{V}_{BT}$ | Nominal water volume in burnt tissue [ml/kg] | I | $120 \varepsilon_B r_{FV}$ | $120 \varepsilon_B r_{FV}$ [30] |
| $\bar{V}_{IT}$ | Nominal water volume in intact tissue [ml/kg] | I | $120(1 - \varepsilon_B r_{FV})$ | $120(1 - \varepsilon_B r_{FV})$ [30] |
| $[\bar{A}_P]$ | Nominal albumin concentration in plasma [g/ml] | I | 0.059 | 0.059 (0.004) [35] |

| Symbol | Description | I/S | Value | Reference |
|---|---|---|---|---|
| $[\bar{A}_{BT}]$ | Nominal albumin concentration in burnt tissue [g/ml] | I | 0.028 | 0.028 [35] |
| $[\bar{A}_{IT}]$ | Nominal albumin concentration in intact tissue [g/ml] | I | 0.028 | 0.028 [35] |
| $\bar{A}_P$ | Nominal albumin content in plasma [g] | I | $[\bar{A}_P]\bar{V}_P$ | - |
| $\bar{A}_{BT}$ | Nominal albumin content in burnt tissue [g] | I | $[\bar{A}_{BT}]\bar{V}_{BT}$ | - |
| $\bar{A}_{IT}$ | Nominal albumin content in intact tissue [g] | I | $[\bar{A}_{IT}]\bar{V}_{IT}$ | - |
| $\bar{J}_C$ | Nominal capillary filtration [ml/kg·h] | S | 1.53 (0.19) | 1.72 [29] |
| $C_O$ | Colloid oncotic pressure constant [mmHg/g·ml] | I | 250 | 250 [35] |
| $\bar{J}_L$ | Nominal total lymphatic flow to plasma [ml/kg·h] | I | 1.07 (0.19) | 1.08 [30] |
| $\bar{J}_{L,BT}$ | Nominal lymphatic flow from burnt tissue to plasma [ml/kg·h] | I | $\bar{J}_L \varepsilon_B r_{FV}$ | $\bar{J}_L \varepsilon_B r_{FV}$ [30] |
| $\bar{J}_{L,IT}$ | Nominal lymphatic flow from intact tissue to plasma [ml/kg·h] | I | $\bar{J}_L(1-\varepsilon_B r_{FV})$ | $\bar{J}_L(1-\varepsilon_B r_{FV})$ [30] |
| $C_L$ | Lymphatic maximal increase coefficient [·] | S | 0.29 (0.21) | - |
| $S_L$ | Lymphatic pressure sensitivity coefficient [1/mmHg] | S | 6.44 | - |
| $\bar{P}_C$ | Nominal hydrostatic capillary pressure [mmHg] | S | 8.0 (0.5) | 6.7 (0.8) [71] |
| $E_C$ | Capillary elastance [mmHg/ml] | S | 0.0139 | 0.0097 [47] |
| $\alpha$ | Tissue electrostatic pressure coefficient [mmHg] | I | 10 | 10 [46] |
| $\gamma$ | Tissue tension pressure coefficient [mmHg] | I | 3.75 | 3.75 [46] |
| $\hat{y}$ | Maximum half-thickness of the extracellular matrix [·] | I | 4 | 4 [46] |
| $\check{y}$ | Minimum half-thickness of the extracellular matrix [·] | I | 1 | 1 [46] |
| $\hat{R}$ | Maximum GAG radius [·] | I | 3.5 | 3.5 [46] |
| $\beta$ | Radius threshold ratio [·] | I | 0.23 | 0.23 [46] |
| $n$ | Hydration response coefficient [·] | I | 8 | 2-8 [46] |
| $\bar{W}_X$ | Nominal hydration level [ml/g] | I | 0.66 | 0.23-0.81 [33], [46] |
| $\bar{J}_{RPF}$ | Nominal renal plasma flow [ml/kg·h] | I | 536 | 536 [72] |
| $\tau_{TGF}$ | Tubuglomerular feedback time constant [1/h] | S | 0.35 (0.33) | - |
| $K_{TGF}$ | Tubuglomerular feedback sensitivity [·] | S | 5.75 | - |
| $\bar{R}_G$ | Nominal glomerular resistance [mmHg/ml/kg·h] | I | 12.34 | - |
| $\lambda_G$ | Glomerular hydrostatic pressure sensitivity [mmHg$^2$/(ml/h·kg)$^2$] | I | 1 | - |
| $P_B$ | Hydrostatic pressure in Bowman's capsules [mmHg] | I | 18 | 18 [53] |
| $K_G$ | Glomerular filtration coefficient [ml/kg·h·mmHg] | S | 9.2 (1.2) | 9.6-12.0 [73] |
| $K_{ADH}$ | Nominal ADH secretion rate [pg/kg·h] | I | 287 | 287 [58], [59] |
| $\lambda_{V_P}$ | ADH sensitivity to plasma volume change [1/ml] | S | 0.0017 (0.0008) | - |
| $\lambda_{[Na^+]}$ | ADH sensitivity to sodium concentration change [l/mEq] | S | 0.087 | - |
| $K_{RR}$ | Maximum collecting duct reabsorption fraction [·] | I | 0.999 | - |
| $[ADH]_{50}$ | ADH concentration corresponding to $\frac{1}{2}K_{RR}$ [pg/ml] | S | 0.0628 | - |
| $[\overline{Na^+}]$ | Nominal plasma sodium concentration [mEq/l] | I | 142 | 142 [74] |
| $\bar{J}_{RR,ADH}$ | Nominal water reabsorption rate in the collecting ducts [·] | S | 0.933 (0.01) | 0.97 [55] |
| $\hat{Q}_{PD}$ | Protein influx post burn [g/h] | S | 72.98 (53.23) | - |
| $\lambda_{PD}$ | Protein influx decay rate [1/h] | S | 10 | - |
| $M_{P_{BT}}$ | Maximum burnt tissue hydrostatic pressure perturbation [mmHg] | S | 38 | 30 [60] |
| $\lambda_{1,P_{BT}}$ | Burnt tissue hydrostatic pressure perturbation slow decay rate [1/h] | S | 5.48 | - |
| $\mu$ | The ratio between slow decay rate to fast decay rate [·] | S | 356 | - |
| $\lambda_{2,P_{BT}}$ | Burnt tissue hydrostatic pressure perturbation fast decay rate [1/h] | I | $8\lambda_{1,P_{BT}}$ | - |
| $K_{1,EV}$ | Nominal tissue evaporation rate [ml/h·m$^2$] | I | 18.48 | 18.48 [66] |
| $\lambda_{1,EV}$ | Evaporation growth rate [1/h] | I | 0.073 | 0.073 [66] |
| $K_{2,EV}$ | Maximum evaporation rate [ml/h·m$^2$] | I | 28.68 | 28.68 [66] |
| $\lambda_{2,EV}$ | Evaporation decay rate [1/h] | I | -0.0052 | -0.0052 [66] |
| $\varepsilon_B$ | Fraction of body surface subject to burn [·] | S | 0.3-0.4 | dataset |
| $S_B$ | Total body surface area [m$^2$] | I | 1 | 1.07+/-0.16 [75] |
| $K_{EX}$ | Maximum exudation rate [ml/h·m$^2$] | I | 25 | 25 [65] |
| $\lambda_{EX}$ | Exudation decay rate [1/h] | I | -0.0038 | -0.0038 [65] |
| $\eta_{EX}$ | Exudate to tissue albumin ratio [·] | S | 0.69 | 0.75 [30] |
| $\bar{\alpha}$ | Nominal albumin to capillary pore radius ratio [·] | S | 0.82 | 0.7-0.9 [47] |
| $k_{PD,BT}$ | Capillary destruction fraction [·] | S | 0.56 | 0.50 [28] |
| $M_{\alpha_{BT}}$ | Maximum pore ratio perturbation in burnt tissue [·] | S | 0.37 | 0.30 [31] |
| $M_{\alpha_{IT}}$ | Maximum pore ratio perturbation in intact tissue [·] | S | 0.16 (0.11) | 0.19 [76] |
| $\lambda_{1,\alpha}$ | Pore ratio slow decay rate [1/h] | S | 0.030 | 0.025 [30] |
| $\lambda_{2,\alpha}$ | Pore ratio fast decay rate [1/h] | I | $\mu\lambda_{1,\alpha}$ | - |
| $M_{P_C}$ | Maximum capillary hydrostatic pressure perturbation [mmHg] | S | 19 (12) | 23+/-5 [64] |
| $\lambda_{1,P_C}$ | Capillary hydrostatic pressure perturbation slow decay rate [1/h] | S | 0.54 (0.08) | - |
| $\lambda_{2,P_C}$ | Capillary hydrostatic pressure perturbation fast decay rate [1/h] | I | $\mu\lambda_{1,P_C}$ | - |

## A.5. Model Fitting via Numerical Optimization

### A.5.1. Classification of Sensitive and Insensitive Subject-Specific Model Parameters

We selected 24 subject-specific parameters in the mathematical model that must be estimated using the datasets, as described in Material and Methods. One dataset [34] had 60 measurements on the average (including 12 PV and 48 UO measurements), while the other dataset [35] had 161 measurements on the average (including 24 PV, 24 UO measurements, 41 lymphatic flow associated with burnt tissues, 50 lymphatic flow associated with intact tissues, 10 plasma albumin concentration, 7 burnt tissue albumin concentration, and 5 intact tissue albumin concentration). The datasets were not deemed too sparse to estimate the subject-specific parameters. But, the amount of measurements in the datasets may not be rich enough to robustly estimate all the 24 subject-specific parameters. Hence, we selected and estimated a subset of sensitive subject-specific parameters on the individual basis while fixing the subject-invariant and insensitive subject-specific parameters to appropriate pre-specified and population-average values. To classify the subject-specific parameters into sensitive and insensitive categories, we examined and compared the degree of inter-individual variability associated with all the subject-specific parameters. First, we determined the population-average parameter values $\bar{\theta}$ by fitting our mathematical model to minimize the cost function in Eq. (37) using the pooled approach [38]:

$$\bar{\theta} = \arg\min_{\theta} \bar{J}(\theta) = \arg\min_{\theta} \sum_{i=1}^{N} \sqrt{\sum_{j=1}^{M_i} \left( \sum_{k=1}^{D_{ij}} \frac{\left|y_{ij}^d(t_k) - y_{ij}(t_k, \theta)\right|}{Y_{ij}} \right)^2}, \qquad (37)$$

where $\theta$ is the vector of subject-specific parameters, $N$ is the number of subjects (=16), $M_i$ is the number of physiological variables measured in the subject $i$ (e.g., $M_i$=2 if PV and UO were measured), $D_{ij}$ is the number of measurements associated with the physiological variable $j$ in the subject $i$ during the initial 48 hours, $y_{ij}^d(t_k)$ is the value of the physiological variable $j$ associated with the subject $i$ measured at time $t_k$, $y_{ij}(t_k, \theta)$ is the value of the same physiological variable at time $t_k$ predicted by the mathematical model for a given $\theta$, and $Y_{ij}$ is the normalization factor for the physiological variable $j$ associated with the subject $i$, which is defined as the range of $y_{ij}^d$ multiplied by $D_{ij}$ (so that the normalized errors associated with all the physiological variables have comparable magnitudes across all subjects). Second, we estimated all the 24 subject-specific parameters $\theta_i$ associated with the subject $i$ by fitting our mathematical model to all the measurements associated with the subject $i$ to minimize the cost function in Eq. (38), using a regularized fitting that minimizes the number of parametric deviations from the population-average values [39], [40]:

$$\theta_i = \arg\min_{\theta} J_i(\theta) = \arg\min_{\theta} \sqrt{\sum_{j=1}^{M_i} \left( \sum_{k=1}^{D_{ij}} \frac{\left|y_{ij}^d(t_k) - y_{ij}(t_k, \theta)\right|}{Y_{ij}} \right)^2} + \lambda_p \sum_{l=1}^{24} \left| \frac{\theta(l) - \bar{\theta}(l)}{\Theta_l} \right|, \qquad (38)$$

where $\lambda_p$ is the regularization weight and $\Theta_l$ is the normalization factor for the $l$-th element $\theta(l)$ of $\theta$, which is defined so that all the elements in $\theta$ are ranged approximately between 0 and 1 (note that such $\Theta_l$ can be estimated by, e.g., solving Eq. (38) with $\lambda_p = 0.05$ and setting $\Theta_l$ as the inter-individual variability of the corresponding element $\theta(l)$). Third, using $\theta_i$ thus estimated from all the 16 subjects, we computed the average normalized deviation of each element in $\theta_i$. Then, we selected those elements of $\theta_i$ associated with deviations thus computed larger than a threshold value as sensitive subject-specific parameters. We set the threshold deviation as 10% based on empiric trial and error, which yielded 12 sensitive subject-specific parameters. Post-hoc parametric sensitivity analysis indicated that the mathematical model was not actually sensitive to one of them, which was thus removed. This exercise yielded a total of 11 sensitive subject-specific parameters (**Table A2**).

**Table A2**: Sensitive subject-specific parameters and their average normalized deviations from population-average values.

| Parameter | $\tau_{TGF}$ | $M_{\alpha_{IT}}$ | $\check{J}_C$ | $\check{J}_{RR,ADH}$ | $\lambda_{V_P}$ | $M_{P_C}$ | $\lambda_{1,P_C}$ | $C_L$ | $\bar{P}_C$ | $\hat{Q}_{PD}$ | $K_G$ |
|---|---|---|---|---|---|---|---|---|---|---|---|
| Deviation [%] | 48.2 | 18.8 | 16.9 | 16.7 | 16.4 | 15.3 | 14 | 13.2 | 11.6 | 11.5 | 10.3 |

### A.5.2. Model Fitting and Estimation of Subject-Specific Model Parameters

We estimated the 11 sensitive subject-specific parameters selected above on the individual basis by fitting our mathematical model to all the available measurements associated with each subject to minimize the cost function in Eq. (39), while fixing the remaining 13 insensitive subject-specific parameters to respective population-average values ($\bar{\theta}$ in Eq. (37)), and as well, fixing the subject-invariant parameters to respective pre-specified values (see **Table A1**):

$$\check{\theta}_i = \arg\min_{\check{\theta}} \check{J}_i(\check{\theta}) = \arg\min_{\check{\theta}} \sqrt{\sum_{j=1}^{M_i} \left( \sum_{k=1}^{D_{ij}} \frac{\left| y_{ij}^d(t_k) - y_{ij}(t_k, \check{\theta}) \right|}{Y_{ij}} \right)^2} \quad (39)$$

where $\check{\theta}$ is the vector of sensitive subject-specific parameters, and $\check{\theta}_i$ is $\check{\theta}$ estimated for the subject $i$.

### A.5.3. Numerical Optimization Details

The complexity and nonlinearity associated with our mathematical model strongly suggest the non-convex nature of the numerical optimization problems in Eq. (37)-(39). To derive robust estimates of parameters from our numerical optimization problems, we used multiple initial conditions and tight parameter bounds as follows. First, we used a multi-start gradient descent method in MATLAB ("globalsearch" in conjunction with "fmincon" commands). We empirically selected user-configurable settings (e.g., the number of initial conditions) so that the numerical optimization could yield accurate parameter estimates when simulated measurements associated with the population-average model (i.e., our mathematical model characterized by $\bar{\theta}$ in Eq. (37)) are inputted to Eq. (39). Second, we enforced tight parameter bounds as constraints in solving the numerical optimization problems to effectively guide the solution into a mechanistically plausible parameter space. We carefully specified many of these bounds by leveraging the prior knowledge on the parameter values (see **Table A1**). It is noted that we intended to also avoid overfitting with these parameter bounds. Since the amount of measurements in our datasets may not be sufficiently large to robustly solve our numerical optimization problems, mechanistically plausible parameter bounds are expected to benefit in preventing overfitting against measurement noises and errors by keeping the parameter estimates from assuming mechanistically illegitimate values.

All in all, our approach to solve the model fitting problem in Eq. (39) by incorporating (i) a small number of sensitive subject-specific parameters, (ii) multi-start gradient descent, and (iii) parameter bounds appeared to be effective: when Eq. (39) was solved using simulated measurements associated with the population-average model repeatedly, NMAE<0.1% was consistently achieved, and all the 11 sensitive subject-specific parameters had very small errors of <3%.

#### FUNDING

This research was supported in part by the U.S. Army SBIR Program [Award No. W81XWH-16-C-0179], the Congressionally Directed Medical Research Programs [Award No. W81XWH-19-1-0322], and the U.S. National Science Foundation CAREER Award [Award No. 1748762].